\documentclass[prl,superscriptaddress,showpacs,floatfix,twocolumn]{revtex4}

\usepackage{amssymb}
\usepackage{amsmath}
\usepackage{graphicx}

\begin{document}

\title{Onset of an Insulating Zero-Plateau Quantum Hall State in Graphene}
\author{E. Shimshoni}
\address{Department of Physics, Bar-Ilan University, Ramat-Gan 52900, Israel}
\address{Department of Mathematics–Physics, University of Haifa at Oranim, Tivon 36006, Israel}
\author{H.A. Fertig}
\address{Department of Physics, Indiana University, Bloomington, Indiana 47405}
\address{Department of Physics, Technion, Haifa 32000, Israel}
\author{G. Venketeswara Pai}
\address{Department of Physics, Technion, Haifa 32000, Israel}
\address{Department of Mathematics–Physics, University of Haifa at Oranim, Tivon 36006, Israel}

\pacs{71.10.Pm, 72.10.Fk, 73.20.-r, 73.23.-b, 73.43.Nq}

\begin{abstract}
We analyze the dissipative conductance of the zero-plateau quantum
Hall state appearing in undoped graphene in strong magnetic fields.
Charge transport in this state is assumed to be carried by a
magnetic domain wall, which forms by hybridization of two
counter--propagating edge states of opposing spin due to
interactions. The resulting non--chiral edge mode is a Luttinger
liquid of parameter $K$, which enters a gapped, perfectly conducting
state below a critical value $K_c\approx 1/2$. Backscattering in
this system involves spin flip, so that interaction with localized
magnetic moments generates a finite resistivity $R_{xx}$ via a
``chiral Kondo effect''. At finite temperatures $T$, $R_{xx}(T)$
exhibits a crossover from metallic to insulating behavior as $K$ is
tuned across a threshold $K_{MI}$. For $T\rightarrow 0$, $R_{xx}$ in
the intermediate regime $K_{MI}<K<K_c$ is finite, but diverges as
$K$ approaches $K_c$. This model provides a natural interpretation
of recent experiments.

\end{abstract}

\date{\today}
\maketitle

\textit{Introduction}-- Graphene, a honeycomb network of carbon
atoms, is perhaps the most remarkable two-dimensional system to be
studied in the last few years \cite{review}.  This system differs
from the standard two dimensional electron gas (2DEG) in
supporting two Dirac points in its fermion spectrum which are at
the Fermi energy when the system is nominally undoped. In the
presence of a magnetic field, the Landau level spectrum differs
from the standard 2DEG in supporting positive and negative energy
(Landau level) states, as well as a zero energy (lowest) Landau
level in each of its two valleys.  In the earliest experiments,
this last property was understood to account for the absence of a
quantized Hall effect at filling factor $\nu=0$
\cite{zheng,zhang}.

In stronger fields, and with higher quality samples, a plateau
does appear to emerge near $\nu=0$ which is likely associated with
a resolution of the spin-splitting of the lowest Landau level
\cite{zhang06,abanin07,ong07}. However, the behavior of the
longitudinal (dissipative) resistance ($R_{xx}$) at this filling
has been difficult to explain: different measurements have
demonstrated that it may either decrease \cite{abanin07} or
increase \cite{ong07} with falling temperature $T$ in samples that
appear to be quite similar \cite{novoselov_pr}. Moreover, in Ref.
\cite{ong07} the $\nu=0$ quantized Hall state exhibits a transition to
an insulator at a critical magnetic field $H_c$, where $R_{xx}(T\rightarrow 0)$
diverges. The scaling of $R_{xx}$
as a function of field $H$ below $H_c$ appears to signify a quantum phase
transition of the Kosterlitz--Thouless (KT)\cite{KT} type.

In this paper, we demonstrate that such a variety of transport
behaviors can be explained quite naturally if the system contains
local magnetic moments near its edge, which may occur due to
chemical passivation of dangling bonds or imperfections in the
lattice \cite{lehtinen04}. Interaction with such magnetic degrees of
freedom provides a mechanism for back--scattering in the primary
channel for charge conduction at $\nu=0$, a magnetic domain wall
(DW) \cite{brey06,abanin06,fertig06} which may carry
current that is unaffected by ordinary, {\it non--magnetic}
impurities. Transport properties of this mode are dramatically
altered by tuning of a Luttinger parameter $K$
\cite{giamarchi_book}, which in this system depends on the details
of the edge potential and on the magnetic field $H$.

\begin{figure}[t]
\includegraphics[scale=.5]{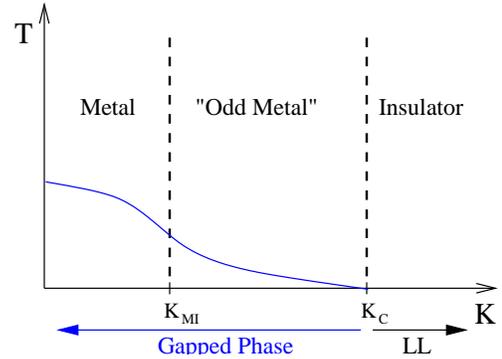}
\caption{(Color online). Phase diagram of the DW coupled to a
magnetic impurity via a fixed Kondo interaction. The solid blue
curve depicts the gap $\Delta_s(K)$. The dashed lines mark the
boundaries between distinct transport regimes (see text).}\label{phases}
\end{figure}

Our main findings are summarized in the phase diagram depicted in
Fig. \ref{phases}. For $K>K_c$, the ideal system is a Luttinger
liquid, with $K$ large enough that a Kondo
impurity renders the system insulating at zero temperature
[$R_{xx} \rightarrow \infty$].
As $K$ is tuned below $K_c$, the impurity-free DW system
undergoes a quantum phase transition
to a perfect conducting state protected by a gap $\Delta_s(K)$
(solid line in Fig. 1). In this case a Kondo impurity generates
a {\it finite} resistance at zero temperature, approaching
this value from {\em below} ($dR_{xx}/dT < 0$).
The unusual thermal behavior of this
``odd metal,''  results from a competition between
two relevant perturbations which separately would render the
system either perfectly conducting or truly insulating.
Finally, for $K<K_{MI}$, a Kondo impurity is an irrelevant
perturbation, yielding more typical metallic behavior,
$dR_{xx}/dT > 0$.
This variety of behavior is consistent with existing
published data \cite{abanin07,ong07}.

\begin{figure}[t]
\includegraphics[scale=.35]{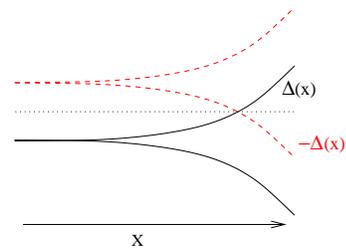}
\caption{(Color online). Non--interacting energy levels near the
edge. Solid black lines are spin up states, dashed red lines are
spin down states. The dotted line denotes the chemical potential
at zero energy.} \label{levels}
\end{figure}

The remarkable edge--mode structure of the DW can already be seen
within a  non-interacting picture. The lowest Landau levels support
an electron-like and a hole-like mode for each spin, which cross
precisely at the Fermi level when Zeeman splitting is introduced
[see Fig. \ref{levels}] \cite{brey06,abanin06,mele05}. Accounting for
interactions within a Hartree-Fock description, the
counter--propagating states admix to form the DW \cite{fertig06}, with
an unusual collective mode. In particular,
due to the spin--charge coupling inherent in a state projected into
a single Landau level \cite{kane}, it supports topological {\em
charged} excitations which can propagate freely along the edge.
In addition, a spin gap may open in the collective excitation spectrum
when interactions are not too strong, with important consequences
for dissipative transport.

A prominent candidate for a dissipation mechanism in this system is
the scattering between counter--propagating edge states
\cite{kane_rev}, which one might expect to be enhanced
by the DW structure which strongly hybridizes forward and backward
propagating states. However, the requirement of spin conservation
forbids backscattering within a single DW by static impurities
\cite{abanin06}. In contrast, scattering from local {\it magnetic}
impurities {\it does} allow backscattering and generates
dissipation. As we show below, the coupling of a DW to a magnetic
impurity leads to an unusual ``chiral'' Kondo effect, since left and
right moving electrons each come in only a single spin flavor. The
resulting resistance at finite $T$ exhibits the variety of behaviors
summarized above, which we discuss below in more detail.

\textit{Model} -- As a first stage we derive an effective model for
a clean graphene sheet, with a straight edge along
the $y$ direction.
Our starting point for the analysis is to view
the system as a two-dimensional (2D) quantum Hall ferromagnet, in
which only the two Landau levels closest to zero energy are
included. These levels carry opposite spin, and
may be grouped together into a spinor whose
long-wavelength degrees of freedom are those of a Heisenberg
ferromagnet, with Hamiltonian \cite{kane,fertig06}
\begin{equation}
H_{2D}^{0}= \int d^2r \left\{{J \over 2} \sum_{\mu}|\partial_{\mu}
{\bf S}({\bf r})|^2+\Delta(x)S_z({\bf r})\right\}.
\label{2Dferr}
\end{equation}
Here $J$ represents an exchange stiffness due to interactions.
$\Delta(x)$ encodes the combined effects of the Zeeman interaction
and the edge potential (dictated by the boundary conditions and electrostatic
environment \cite{silvestrov}): it increases monotonically from the bulk
(negative) Zeeman energy for $x$ well inside the bulk to a positive
value at the edge, leading to the energy crossing  in the
non-interacting spectrum illustrated in Fig. \ref{levels}. Initially
treating the system classically, we parameterize the spin field by
${\bf S}({\bf r}) = S {\bf \Omega}({\bf r})$, with $S=1/2$ and ${\bf
\Omega}= [\sin\psi({\bf r})\cos\chi({\bf
r}),\sin\psi({\bf r})\sin\chi({\bf r}),\cos\psi({\bf
r})]$. As shown in \cite{fertig06}, the form of $\Delta(x)$ in
the above energy functional dictates a minimum energy configuration
where $\chi$ is spatially constant, and $\psi({\bf
r})=\psi_0(x)$ exhibits a non-trivial topology: $\psi_0(x
\rightarrow -\infty) \rightarrow 0$, $\psi_0(x \rightarrow {\rm
edge}) \rightarrow \pi$. A domain wall (DW) is thus formed parallel
to the edge. The ground state energy is independent of $\chi$,
implying a broken symmetry and an associated collective mode
propagating with a wavevector $k_y$ along the DW \cite{falko}.

The 2D semiclassical theory may be projected into its low-energy
subspace, containing the one-dimensional (1D) mode.  This theory is
identical in form to a semiclassical treatment of an XXZ antiferromagnetic
spin-1/2 chain;
an appropriate choice of coefficients for the latter allows the
models to coincide
at the semiclassical level.  As the spin chain
incorporates the spin-1/2 nature of the system,
we work with this model, with lattice spacing
set by the
magnetic length $\ell=\sqrt{\hbar c/eH}$:
\begin{equation}
H_{DW}=J_{xy}\sum_{j}\left[ S^x_{j} S^x_{j+1} + S^y_{j} S^y_{j+1} \right] +
J_z\sum_{j}S^z_{j} S^z_{j+1}\; . \label{XXZ}
\end{equation}
Here $J_{xy}$ is related to the exchange energy $J$ via
$J_{xy}=JS^2{\cal N}$, with ${\cal N}=\sum_x \sin^2
\psi_0(x)$
($\psi_0(x)$ the classical DW texture), and $J_z$ depends on
the edge potential: $J_{z}={1 \over {{\cal N}2S}} \sum_x \Delta(x)
\sin^2 \psi_0(x)[1-\cos \psi_0(x)]$.
Using a standard Bosonization scheme for the spin operators in the
continuum limit \cite{giamarchi_book},
\begin{eqnarray}
S_\pm(y)=\frac{e^{\mp
i\theta}}{\sqrt{2\pi\alpha}}[(-)^y+\cos(2\phi)]\nonumber\\
S_z(y)=-\frac{\partial_y
\phi}{\pi}+\frac{(-)^y}{\pi\alpha}\cos(2\phi)\; ,\label{boson}
\end{eqnarray}
we map $H_{DW}$ to a (single--flavored) Luttinger model with a
sine--Gordon correction. In units where $\hbar=1$,
\begin{equation}
H_{DW}=\int \frac{dy}{2 \pi} \left\{ uK (\pi\Pi) ^2+\frac{u}{K}
(\partial_y \phi)^2 -\frac{J_z} {\pi \alpha^2} \cos(4\phi)\right\}
\label{LL}
\end{equation}
where the Boson field $\phi(y)$ and $\Pi(y)={\partial_y\theta\over
\pi}$ are canonically conjugate.
The spin--wave velocity $u$ and Luttinger parameter $K$ are related
to the XXZ parameters by
\begin{equation}
uK=J_{xy}\; ,\quad \frac{u}{K}=J_{xy}+\frac{4J_z}{\pi}\; ,
\label{LLdef}
\end{equation}
and the short distance cutoff $\alpha\sim\ell$.

The effective model for the ideal DW system [Eq. (\ref{LL})]
exhibits a quantum phase transition upon tuning of
the parameter $K$ across a critical point $K_c$, from a Luttinger liquid (LL) phase at $K>K_c$
[where the $\cos(4\phi)$ term is irrelevant] to an ordered [gapped] phase below $K_c$.
In terms of the spin fields, the latter phase is characterized
by ordering of the field $S_z$, and the
opening of a gap $\Delta_s$ to spin-flip excitations. The
critical behavior near $K_c$ can be derived from a perturbative
renormalization group (RG) analysis \cite{giamarchi_book}, yielding
a $KT$--transition at $K_c\sim 1/2$. As $K$ approaches $K_c$ from below,
the gap tends to vanish according to the scaling law
\begin{equation}
\Delta_s(K) \simeq
\frac{u}{\alpha}\exp\left\{{-\frac{C}{\sqrt{K_c-K}}}\right\}
\label{Delta_s}
\end{equation}
with $C$ a constant of order unity [see Fig. \ref{phases}].

We now assume that a local moment at the origin (represented by a
spin-1/2 operator $\vec{\sigma}$) couples to the original 2D spin
system via an exchange interaction $\tilde{J} {\vec S} ({\bf
r}=0)\cdot {\vec \sigma}$. In order to project this into the
low--energy subspace, we express it in a rotated basis where the
local moment is coupled to fluctuations about the ground state DW
configuration. This yields an effective Kondo coupling to the 1D
spin chain operators:
\begin{equation}
H_K=\frac{J_K^{xy}}{2} (S_+(0)\sigma_- + S_-(0)\sigma_+) + J_K^z
S_z(0)\sigma_z\; . \label{kondo}
\end{equation}
The resulting Hamiltonian for the effective 1D system then takes the
form $H=H_0+H_{K}$, with
\begin{equation}
H_0=H_{DW}+\tilde{E}_z\sigma_z\; .\label{H0}
\end{equation}
Note that $\tilde{E}_z$ is an effective Zeeman energy of the
localized spin, associated with a {\it local} magnetic field which
depends on its position within the width of the DW.

\textit{Transport} -- To derive the transport properties of the
system, we first express the charge current operator in terms of the
DW fields. Using the standard definition $j_e=-{\delta
H_{DW}\over\delta A}$, with $A(y,t)$ a vector potential, we find
\begin{equation}
j_e(y,t)=2eJ_z S_z(y,t)\; .\label{je}
\end{equation}
This remarkable relation between spin and charge current encodes a
unique property of the DW: in the absence of spin-flip processes
(i.e. for $J_K^{xy}=0$), current is conserved and the DW behaves as
a perfect conducting channel. The distinction between its two phases
becomes apparent when spin impurities are added: in the gapped phase,
spin-flip and hence back-scattering excitations are suppressed for $T<\Delta_s$.
Its `perfect conduction' is hence more robust.
Indeed, $H_{DW}$ [Eq. (\ref{LL})] is
analogous to a low--energy model of a quasi 1D {\em superconducting}
system, where two parallel wires are Josephson coupled. The field
$\phi$ describes the relative superconducting phase, $u/K$ the
superfluid stiffness, $uK$ the charging energy \cite{1dSC_rev}, and
the last term in $H_{DW}$ is the inter-wire Josephson coupling.
There as well, a local perturbation analogous to
$H_K$ [Eq. (\ref{kondo})] is required to generate finite dissipation
via a phase--slip mechanism. This occurs when a Josephson vortex
(i.e., out-of-phase current configuration) penetrates into the bulk
of the double-wire system. In the gapped phase, where Josephson
coupling is relevant, such processes are suppressed beyond the
Josephson length ($\sim 1/\Delta_s$).

We next consider a finite but small Kondo interaction and
$T>\Delta_s$, and calculate the dissipative part of the electric
conductance to leading order in $H_K$:
\begin{equation}
\delta G=\lim_{\omega\rightarrow
0}\frac{-1}{L\omega}\int_{-L/2}^{L/2}dy \Im
\left\{\chi(y,y^\prime;\omega)\right\}\; , \label{Kubo}
\end{equation}
where $\chi=\langle j_e(y);j_e(y^\prime)\rangle^0_{\omega}$ is the
retarded correlation function evaluated with respect to $H_{0}$
(see, e.g., \cite{giamarchi_book}), and $L$ the length of the
system, assumed finite. Note that $\delta G$ is a {\it negative}
correction to the ideal conductance, so that $|\delta G|$ is
proportional to the backscattering rate and consequently to the
longitudinal resistance $R_{xx}$ \cite{Gdef}. Using the bosonized
representation Eqs. (\ref{boson}),(\ref{LL}) and the expression for
the local spin correlator (for $t>0$)
\begin{equation}
\langle
\sigma_+(t)\sigma_-(0)\rangle^0=e^{i\tilde{E}_zt}f\left(\frac{\tilde{E}_z}{T}\right)\;
,\quad f(z)\equiv \frac{1}{e^z+1}
\end{equation}
we obtain the $T$--dependent conductance
\begin{equation}
\delta G(T)\approx -\frac{e^2}{h}g_K^2 \left(\frac{\pi\alpha
T}{u}\right)^{\kappa-2}\mathcal{G}\left(\frac{\tilde{E}_z}{T},\frac{\alpha
T}{u}\right)\; .\label{deltaG}
\end{equation}
Here $g_K\propto J_K^{xy}$ is dimensionless, $\kappa=1/2K$ and
\begin{eqnarray}
\mathcal{G}(z,\epsilon)&\equiv &
\sin\left(\frac{\pi\kappa}{2}\right)\int_\epsilon^\infty
dt\frac{t\cos(zt)}{[\sinh(\pi t)]^\kappa}  \nonumber\\
&-& 2m(z)\cos\left(\frac{\pi\kappa}{2}\right)\int_\epsilon^\infty
dt\frac{t\sin(zt)}{[\sinh(\pi t)]^\kappa}\label{Gdim}
\end{eqnarray}
with $m(z)=[f(z)-f(-z)]$ the local spin magnetization.

In the finite $T$ regime where $T\gg \tilde{E}_z$, the leading
$T$--dependence of $\delta G(T)$ inferred from Eqs. (\ref{deltaG}),
(\ref{Gdim}) is a power--law, which changes sign at $\kappa=2$:
\begin{equation}
\delta G(T)\sim -T^{\kappa-2}\; .\label{GvsT}
\end{equation}
The conductance therefore exhibits a crossover from a metallic
behavior ($dG/dT<0$) for $\kappa>2$ to an insulating behavior
($dG/dT>0$) for $\kappa<2$. This crossover
would be a true metal--insulator (MI) as $K$ is tuned
through $K_{MI} \approx 1/4$, as depicted
in Fig. (\ref{phases}), if not for the existence of the gap.
Note that $K_{MI}$ is perturbatively
renormalized to values below 1/4 by $J_K^{xy}$ and $J_K^{z}$, so
that one always finds $K_{MI}<K_c$.
The MI crossover then always occurs within the gapped
phase. Thus the power--law behavior of $G(T)$ is
restricted to finite $T\gg \Delta_s$.

We now focus on the intermediate regime $K_{MI}<K<K_c$. Assuming
further that $\tilde{E}_z\sim 0$, we consider the low $T$ limit
$T\ll\Delta_s$ where the field $\phi$ is ordered, while $\theta$ is
strongly fluctuating. The power--law decay of the correlation
function $\langle e^{i\theta(t)}e^{-i\theta(0)}\rangle^0$
(characteristic of the LL) is modified by a rapidly oscillating
exponential factor:
\begin{equation}
\langle e^{i\theta(t)}e^{-i\theta(0)}\rangle^0\approx
\left(\frac{\alpha}{ut}\right)^\kappa
\exp\left\{-\frac{\kappa\Delta_s
t^2}{\tau_\phi}+i\kappa\left(\frac{\pi}{2}+\frac{t}{\tau_\phi}\right)\right\}
\label{lowTcorr}
\end{equation}
where the decoherence rate $\tau_\phi^{-1}\equiv\Delta_s^2L/2\pi u$
corresponds to a `condensation energy' in the superconductor
analogue. The resulting $\delta G$ saturates as $T\rightarrow 0$ and
becomes
\begin{equation}
\delta G\sim -\tau_\phi^{2-\kappa}\; .
\end{equation}
It then follows from the definition of $\tau_\phi$ and
Eq. (\ref{Delta_s}) that the resistance {\it diverges} when $K$
approaches $K_c$ as
\begin{equation}
R_{xx}\sim -\delta G\sim
\exp\left\{\frac{2C^\prime}{\sqrt{K_c-K}}\right\}\; .
\label{RvsK}
\end{equation}
Here $C^\prime=C(2-\kappa)$; for $\kappa\sim 1/2K_c\sim 1$,
$C^\prime\sim C$.

We finally comment on the possible relation of our results to the
experimentally observed $R_{xx}$ in the
zero--plateau state. Provided $K$ is tunable by a magnetic field $H$, the system
is shown to exhibit diverse transport properties, which could explain the
apparent discrepancy between the
data of Refs. \cite{abanin07} and  \cite{ong07}.
Furthermore, we find a quantum critical point associated with the
closing of a spin--gap, manifested by a divergence of
$R_{xx}(T\rightarrow 0)$ according to a scaling law characteristic of
a KT--transition (in $1+1$--dimensions). The latter is remarkably
reminiscent of the peculiar field-tuned transition to an insulator observed
in \cite{ong07}.
A rough estimate of $K$ in terms of physical parameters yields $K\sim \sqrt{\frac{e^2}{\epsilon\ell \Delta_{av}}}$,
with $\Delta_{av}$ an average edge potential strongly dependent on sample details.
In particular, when $\Delta_{av}$ is dominated by electrostatic effects induced by the gate voltage
\cite{silvestrov}, it is typically larger than the exchange energy $\frac{e^2}{\epsilon\ell}$.
This would yield $K<1$ and monotonically increasing with $H$, as assumed by our theory.

Another assumption of our theory which requires justification is the
neglect of $\tilde{E}_z$. A finite $\tilde{E}_z$ introduces a cutoff
which competes with the gap $\Delta_s$, and a finite magnetization
of the local moment which reduces its contribution to
backscattering. We note, however, that the realistic system
presumably contains a multitude of spin impurities with randomly
distributed local parameters. The dissipative resistance is then
dominated by the moments with minimal $\tilde{E}_z$, and our
analysis applies as long as it is smaller than all other energy
scales.

To summarize, we study a model for charge transport carried by
fluctuations of a DW propagating along the edge of a graphene sample
in the $\nu=0$ quantized Hall state. Backscattering induced by
localized magnetic impurities at the edge provides a dissipation
mechanism. Its competition with a spin--gap generated in the ideal DW
may possibly explain the rich variety of
conductance properties observed in recent experiments.

\bigskip
\begin{acknowledgments}
We gratefully acknowledge discussions with L. Brey, K.S. Novoselov
and N. P. Ong. This project was supported in part by the US NSF
under Grant No. DMR-0704033 (H.A.F.) and the German--Israeli
Foundation for Scientific Research and Development (E.S. and G.V.P.).
\end{acknowledgments}


\begin{thebibliography}{99}
\bibitem{Gdef} Note that a finite size perfect conductor may also have resistance
associated with its leads \cite{giamarchi_book,abanin06}.
\end{thebibliography}
\end{document}